# Indian Stock Market Prediction using Augmented Financial Intelligence ML

Anishka Chauhan[1], Pratham Mayur[2], Yeshwanth Sai Gokarakonda[3],
Pooriya Jamie[4], Naman Mehrotra[5]

[1]University College London, [2]National Institute of Technology Patna, [3]SRM University AP, [4]Amirkabir University of Technology, Tehran Polytechnic, [5]IIT Kanpur

anishka.chauhan242@gmail.com, prathamnitp21@gmail.com, yeshwanthsai31@gmail.com, pooriya.Jamie@gmail.com, namanm21@iitk.ac.in

## Abstract

This paper presents price prediction models using Machine Learning algorithms augmented with Superforecasters' predictions, aimed at enhancing investment decisions. Five Machine Learning models are built, including Bidirectional LSTM, ARIMA, a combination of CNN and LSTM, GRU, and a model built using LSTM and GRU algorithms. The models are evaluated using the Mean Absolute Error (MAE) to determine their predictive accuracy. Additionally, the paper suggests incorporating human intelligence by identifying "Superforecasters" and tracking their predictions to anticipate unpredictable shifts or changes in stock prices (Mihov et al, 2022). To collect user predictions and identify Superforecasters, a user-friendly website has been developed. The predictions made by these users can further enhance the accuracy of stock price predictions when combined with Machine Learning and Natural Language Processing (NLP) techniques.

Predicting the price of any commodity can be a significant task but predicting the price of a stock in the stock market deals with much more uncertainty. This is mainly due to the market's volatile and disruptive nature. Prices can touch their 52-week high on one day and the other, we might witness a huge fall in price. Market dynamics plays a major role in deciding the price of stocks but there can be an enormous number of reasons behind the price play for this market. Recognising the limited knowledge and exposure to stocks among certain investors, this paper proposes price prediction models using Machine Learning algorithms. These models assist investors who have been hesitant to engage in stock market investments due to the lack of knowledge and experience. In this work, five Machine learning models are built using Bidirectional LSTM, ARIMA, a combination of CNN and LSTM, GRU and the last one is built using LSTM and GRU algorithms. Later these models are assessed using MAE scores to find which model is predicting with the highest accuracy. In addition to this, this paper also suggests the use of human intelligence to closely predict the shift in price patterns in the stock market The main goal is to identify 'Superforecasters' and track their predictions to anticipate unpredictable shifts or changes in stock prices. To collect this data from people and to identify super forecasters, we have built a user-friendly website using which people can submit their predictions about a certain stock. By leveraging the combined power of Machine Learning and the Human Intelligence, predictive accuracy can be significantly increased.

## Abbreviations:

ML: Machine Learning
AI: Artificial Intelligence
LSTM: Long Short-Term Memory
ARIMA: Autoregressive Integrated Moving Average
CNN: Convolutional Neural Network
GRU: Gated Recurrent Unit
NLP: Natural Language Processing
ANN: Artificial Neural Networks
SVM: Support Vector Machine
BSE: Bombay Stock Exchange

NIFTY: National Stock Exchange FIFTY
MAE: Mean Absolute Error
MSE: Mean Squared Error
BiLSTM: Bidirectional Long Short-Term Memory

# 1. INTRODUCTION

The application of machine learning (ML) algorithms in financial market predictions has been a topic of interest in recent years. The objective of this paper is to provide an overview of the current state of financial intelligence in India and propose a framework for augmenting it using emerging technologies such as Machine Learning and Natural Language Processing (NLP). Several studies have explored this area demonstrating the potential of ML in enhancing financial intelligence.

By identifying differences between AI 1.0 (Machine Intelligence), AI 2.0 (Intelligence Amplification) and AI 3.0 (Augmented Intelligence), a research work by Mihov et al, 2022 explored the interactions between machine and human intelligence. Their work suggests the use of 'Superforecasters' (individuals that tend to make better predictions than the general public) with Machine Intelligence in applications like financial forecasting as due to insufficient training data and low noise ratio, Machine Intelligence systems often tend to underperform. This paper is an application of Mihov et al, 2022's work.

As part of the literature review, we have referred to the paper by Payal Soni et al 2022 as it provides a detailed analysis of the techniques employed in predicting stock prices and the challenges associated with them. We also draw upon other relevant literature to support our proposed framework and highlight its potential benefits for the Indian financial sector.

A study by Huang, J., Chai, J. & Cho, S. (2020) demonstrated the use of Deep Learning algorithms in predicting stock market movements. The study found that Deep Learning algorithms outperformed traditional statistical methods, indicating the potential of these algorithms in financial market predictions.

According to Deshmukh and Saratkar (2019), the forecasting of stock returns has become a significant field of research in recent decades. Initially, researchers attempted to establish a linear relationship between macroeconomic variables and stock returns. However, the discovery of nonlinearity in stock market index returns has led to the emergence of literature on nonlinear statistical modelling. Many of these studies require the specification of a nonlinear model before estimation. Given the noisy, uncertain, chaotic, and nonlinear nature of stock market returns, Deshmukh and Saratkar (2019) argue that artificial neural networks (ANN) have evolved as a better technique for accurately capturing the structural relationship between a stock's performance and its determinant factors compared to other statistical techniques. The authors also note that different studies utilise different sets of input variables to predict stock returns, even when predicting the same set of stock return data, highlighting the ongoing exploration and diversity in approaches to identifying the most effective variables for predicting stock returns.

Another ML algorithm Support Vector Machine has been a popular choice for curating price-prediction models. A study by Patel et al. (2015) confirms this as it uses Support Vector Machine (SVM) to predict the Bombay Stock Exchange (BSE) and found that SVM outperformed other traditional methods in terms of accuracy and reliability. This study highlighted the potential of ML in predicting stock market trends in India.

Mehtab and Sen's study on "Stock Price Prediction Using Convolutional Neural Networks on a Multivariate Time Series" proposes a hybrid approach for stock price prediction in the Indian financial market using machine learning and deep learning-based methods. The authors address the debate surrounding the predictability of stock prices by proponents of the Efficient Market Hypothesis who claim that stock prices cannot be predicted, and those who have shown that, if correctly modelled, stock prices can be predicted with a fairly reasonable degree of accuracy.

The authors exploit the power of CNN in forecasting the future NIFTY index values using three approaches which differ in the number of variables used in forecasting, the number of sub-models used in the overall models, and the size of the input data for training the models. The results demonstrate that the CNN-based multivariate forecasting model is the most effective and accurate in predicting the movement of NIFTY index values with a weekly forecast horizon. This study provides valuable insights into the use of machine learning and deep learning-based models for stock price prediction in the Indian financial market, which can be useful for developing augmented financial intelligence tools and strategies.

Overall, Mehtab and Sen's study offers a compelling approach to stock price prediction using machine learning and deep learning-based methods. The authors' use of the NIFTY 50 index values of the NSE of India provides a valuable dataset for the development and evaluation of predictive models. The study's findings demonstrate the potential of CNN-based multivariate forecasting models in predicting stock price movements with high accuracy and provide useful insights for the development of augmented financial intelligence tools and strategies in the Indian financial market.

The paper "Machine Learning Approaches in Stock Price Prediction: A Systematic Review" by Payal Soni et al provides a comprehensive literature review of various techniques used in predicting stock prices. The authors explore traditional machine learning techniques, deep learning, neural networks, time series analysis, and graph-based approaches. They also discuss the challenges associated with each approach and highlight future research directions.

["Stock Market Prediction via Deep Learning Techniques: A Survey"](#) is a literature review that provides a comprehensive overview of the latest deep learning models and techniques for stock market prediction. The authors motivate the need for machine learning techniques in stock market prediction, and present four distinct tasks in this field: stock movement prediction, stock price prediction, portfolio management, and trading strategies.

The survey categorizes the different deep learning models used in stock market prediction using a novel taxonomy, which includes Recurrent Neural Networks (RNNs), Long Short-Term Memory (LSTM), Gated Recurrent Units (GRUs), Graph Neural Networks (GNNs), Convolutional Neural Networks (CNNs), Transformer-based models, and Reinforcement learning (RL) models. The authors analyse 94 papers from top conferences and summarize the datasets, evaluation techniques, and model inputs used in these studies.

The survey also identifies unresolved challenges and potential future research directions in deep learning-based stock market prediction. The authors point out that while deep learning models have shown promising results in stock market prediction, there are still challenges related to data quality, model interpretability, and generalization across different market conditions. They suggest future research directions such as incorporating external data sources, developing more explainable models, and studying the impact of different market conditions on model performance.

Overall, "Stock Market Prediction via Deep Learning Techniques: A Survey" provides a valuable resource for researchers, practitioners, and educators interested in the latest advancements in deep learning techniques for stock market prediction. The survey's novel taxonomy and comprehensive analysis of the datasets and evaluation techniques used in this field can serve as a guide for future research in this area.

In conclusion, the literature suggests that ML algorithms hold significant potential in predicting financial market trends. In the context of India, studies have demonstrated the effectiveness of various ML algorithms in predicting stock market trends, thus enhancing financial intelligence. However, more research is needed to further explore the potential of ML in augmenting financial intelligence in India.

## 2.     CHALLENGES FACED

In India, the percentage of the population that invests in stock markets is fairly low when compared to the USA and Europe. To put this into perspective, let us look at some numbers published by choiceindia.com in June 2023. In the USA, 55% of their citizens invest in the stock market, in the United Kingdom this number is around 33%. But in the case of India, only 3% of its population invests in this market. The biggest reason for this investor gap is the lack of information. Most Indians still live with the stigma that stock markets are not for everyone, only the people who track markets on an everyday basis and keep track of company happenings can invest in this asset. To address this issue and fill in the information gap for people who want to invest in stocks but have no time to manually keep track of stock price, management decisions, quarter numbers, etc. we wanted to curate a financial big data platform that can hold live as well as historical data related to Indian stocks, companies, and products.

The motive was to build a single window platform for providing all the finance-related information required by any individual who wants to acquaint themself with the factors that might impact the price of a particular stock or the market as a whole, before investing in the Indian stock market. For this reason, we did not want to solely rely on the macro data like price, volume, etc. of the stock but we wanted to build a platform that keeps an account of the company-related news, mergers and acquisitions, quarter numbers of the companies, etc. as all these can impact the value of the shares to a great extent.

Along with these factors we wanted to include another way of understanding the price behaviour of stocks and that is by identifying and keeping track of predictions made by 'Super Forecasters'. We extracted the Indian Macro Data from the internet from sites like finance.yahoo.com and investing.com. For collecting nominal data, i.e., predictions made in forums, comments, blogs, news, etc. we had to try different ways. We tried extracting this data using Manual web scrapping, which we were able to achieve but since we wanted to make a big data platform with live data inflowing at all times, manual scrapping was not a sustainable option, in addition to this it was a tedious task to scrape 5-6 websites for different comments and forums of multiple authors to recognize super forecasters.

As a solution to this problem, we came up with automated web scraping. Now the issue with automated scraping is that most websites, especially financial websites have a strong firewall in place that does not allow any automated data scraping activity. Now our next plan of action was to use a third-party web proxy service, which can be used to extract this data even with the firewall in place. On further research, we found out that all of these third-party web proxy services are paid and since the research did not have any funds in place, we could not go ahead with this way of data procurement.

Due to all these reasons stated above, we faced a huge problem of data insufficiency while working on this research. We did not have access to any financial forums or news website data for us to find out potential super forecasters. To conclude, financial data procurement is a tedious task to carry out in Indian markets. This data is not readily available in the case of India as it is in the case of the USA and the UK.

To solve this problem of data insufficiency, we came up with the idea of collecting this data of stock predictions made by humans, first-hand by making a website. Since we could not arrange for any data that could enable our research to find potential super forecasters, we designed and hosted a website that lets users sign in and then make predictions about the price of a stock in the near future. According to how close these predictions are made; users will be ranked on a leader board. We can identify potential super forecasters if a user remains on this leader board for a considerable amount of time. This way we can include human intelligence in making predictions for price along with using Machine Learning algorithms for this work. In the following section, we will be discussing the Price Prediction models that we have built using different machine learning models. Here is the link to the site we have curated for solving the problem of data insufficiency and to identify super forecasters: https://augmentedfinancial.wixsite.com/stockmarketprices

# 3. PROPOSED WORK

This study aimed to understand how the Indian Stock Market prices fluctuate and how can we predict the price of the market to help the Indian investor make informed decisions. As a solution to this problem, to predict the price of Indian stocks, we have successfully built three different price prediction models that use Machine Learning to predict the price of a stock by giving the historical data of that stock to the algorithm. Through this study, we want to understand how robust these models are. To get an idea of the entire Stock Market of India, we are using the NIFTY 50 stock data to compare the effectiveness of these three models.

During the course of this study, we have worked on curating five price-prediction models. The first one is built using the bidirectional Long Short-Term Memory (LSTM) ML algorithm. The second model uses ARIMA to predict the price of NIFTY 50 stock, the third one is built using the combination of LSTM and CNN (Convolutional Neural Network), the fourth one uses GRU algorithm, and the last one is built using the combination of LSTM and GRU.

## 3.1 Bidirectional LSTM (BiLSTM)

In recent years, the Bidirectional LSTM architecture has gained prominence as a compelling extension to the standard LSTM. This architecture augments the LSTM cell by allowing information flow not only in the forward temporal direction but also in the reverse direction. By capturing dependencies in both directions, the Bidirectional LSTM can uncover contextual cues that may remain hidden in unidirectional models. This innovation has found applications in diverse fields including natural language processing, speech recognition, and, notably, time series analysis.

In this paper, we propose a novel neural network architecture that leverages the power of Bidirectional LSTM layers for the task of time series regression. Our architecture aims to enhance the predictive capabilities of traditional LSTM-based models by effectively capturing intricate temporal relationships present in sequential data. We combine the Bidirectional LSTM layer with a Dense layer to achieve accurate predictions of continuous numerical values.

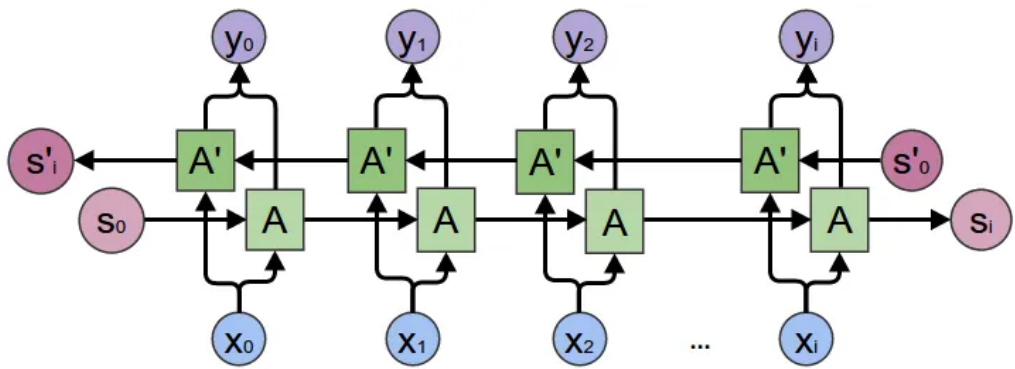

**Figure 1:** BiLSTM Architecture

### 3.1.1 Model Initialization

A sequential model is initialized using the Keras framework's `Sequential` class. This model allows for the construction of a linear stack of layers, facilitating the definition of deep learning architecture.

### 3.1.2 Bidirectional LSTM Layer

To capture the temporal dependencies and patterns within the input data, a bidirectional Long Short-Term Memory (LSTM) layer is employed. The LSTM layer consists of 50 units and uses the rectified linear unit (ReLU) activation function. The bidirectional nature of this layer allows it to process the input data in both the forward and backward directions, enabling a comprehensive understanding of the temporal context. The `input_shape` parameter is specified as `(None, 1, time_step)`, indicating that the input can have any number of time steps.

### 3.1.3 Dense Layer

Following the bidirectional LSTM layer, a dense layer with a single node is added to the model. This layer is responsible for producing the output of the model. The use of a single-node dense layer suggests that the regression task aims to predict a single continuous value as the output.

### 3.1.4 Compilation

To prepare the model for training, it is compiled using the Adam optimizer. The Adam optimizer is a popular choice for deep learning tasks due to its adaptive learning rate and efficient optimization capabilities. The mean squared error (MSE) loss function is employed, which measures the discrepancy between the predicted and true values. By minimizing this loss function, the model aims to learn to accurately predict the target values.

### 3.1.5 Training

The model is trained using the `fit` function, which applies the backpropagation algorithm to update the model's weights iteratively. The training dataset, denoted as `trainX` and `trainY`, represents the input sequences and their corresponding target values, respectively. The training process consists of 50 epochs, indicating that the entire training dataset is traversed 50 times during training. A batch size of 1 is chosen, meaning that each sample is processed individually, allowing for fine-grained updates of the model's parameters.

By utilizing a bidirectional LSTM layer and a single-node dense layer, the model learns to capture and understand the temporal dependencies within the input data, ultimately predicting a continuous value for the given regression task. The training process, consisting of multiple epochs and fine-grained updates with a batch size of 1, enables the model to optimize its parameters and improve its predictive performance.

## 3.2 ARIMA Model

This research also explores the process of building and optimizing ARIMA (AutoRegressive Integrated Moving Average) model parameters for time series forecasting. The study involves experimenting with different combinations of p and q parameters to identify the most accurate model. The Mean Absolute Error (MAE) is employed as a performance metric to evaluate the accuracy of each model. The results highlight the significance of selecting appropriate parameters for achieving better forecasting accuracy in time series analysis.

### 3.2.1 Introduction

Time series forecasting is a critical task in various domains such as finance, economics, and environmental science. ARIMA models have proven to be effective in capturing the temporal patterns of time series data. The objective of this study is to determine the optimal values of the ARIMA model parameters p (order of the autoregressive component) and q (order of the moving average component) to improve the accuracy of forecasting.

### 3.2.2 Background and Related Work

ARIMA models are widely used for time series analysis due to their ability to handle trends and seasonality. Previous research has explored different methodologies for parameter selection, including grid search and optimization techniques. However, in this study, we utilize a nested loop approach to systematically test various combinations of p and q values.

### 3.2.3 Methodology

We consider a time series dataset y as the target variable for forecasting. To find the optimal p and q parameters, we employ a nested loop approach. The p_params and q_params are predefined lists of candidate values for p and q, respectively. For each combination of p and q, an ARIMA model with the order (p, 0, q) is fitted to the data using the ARIMA function. The time taken for model training is recorded using the time.time() function.

### 3.2.4 Results

The results of the experiment are presented, showcasing the performance of each ARIMA model in terms of training time and forecasting accuracy. The MAE is calculated by comparing the predicted values y_pred obtained from each model with the actual values of the time series y. The output displays the trained ARIMA model's order (p, d, q) and the elapsed time for model training in seconds. Subsequently, the corresponding MAE for each model is printed.

### 3.2.5 Discussion

The experimental results demonstrate that the selection of p and q significantly affects the forecasting accuracy of the ARIMA model. Models with larger values of p and q may lead to overfitting, while smaller values may underfit the data.
The analysis of the MAE values reveals the best-performing combination of p and q parameters, which yields the most accurate forecasting results for the given time series dataset.

## 3.3 LSTM and CNN

Sequence prediction, a cornerstone of various applications including speech recognition, financial forecasting, and natural language processing, involves deciphering the underlying temporal patterns within data sequences. This task carries immense significance as it empowers decision-makers with the ability to anticipate future trends, adapt strategies, and make informed choices. In the realm of deep learning, Recurrent Neural Networks (RNNs) and Convolutional Neural Networks (CNNs) have demonstrated their prowess in handling sequential and grid-like data respectively, catalysing advancements in sequence modelling.

RNNs excel at capturing temporal dependencies by leveraging their recurrent connections, making them a natural fit for tasks involving sequences. Conversely, CNNs have proven to be highly effective in extracting spatial features through their convolutional layers, primarily designed for tasks like image recognition. In this research, we propose a novel architecture that converges the strengths of both paradigms—Temporal Convolutional layers and Long Short-Term Memory (LSTM) networks. By doing so, we aim to harness the benefits of temporal hierarchies inherent in sequences while exploiting the ability of LSTM networks to capture long-term dependencies.

The architecture amalgamates Temporal Convolutional layers to capture local temporal features within sequences, enhancing the model's ability to detect significant patterns. The integration of LSTM layers follows, allowing for the model to encode temporal dependencies while retaining computational efficiency. Furthermore, to mitigate overfitting risks inherent in complex architectures, we incorporate dropout regularization in the form of Conv1D layers.

The primary objective of this study is to present a comprehensive analysis of the proposed CNN-LSTM hybrid architecture for sequence prediction tasks. We explore the architecture's effectiveness in capturing intricate temporal dependencies, showcase its predictive accuracy, and underline its potential to enhance sequence prediction capabilities.

In the subsequent sections, we delineate the architectural design, detail the experimental methodology, present empirical results on real-world data, and conclude by discussing the implications of the findings. By innovatively fusing temporal convolution and recurrent memory mechanisms, this research contributes to the growing body of work aimed at creating more robust and effective neural network architectures for sequence modelling.

### 3.3.1 Model Initialization
A sequential model is initialized using the `Sequential` class, which allows for the construction of a linear stack of layers. This model will be used to learn patterns and make predictions on sequential data.

### 3.3.2 TimeDistributed Conv1D Layer
To capture temporal dependencies within the sequential data, a TimeDistributed layer is added to the model. This layer contains a Conv1D layer with 64 filters and a kernel size of 1, which enables the extraction of local features. The rectified linear unit (ReLU) activation function is applied to introduce non-linearity into the model. The `input_shape` parameter is set to `(None, 1, time_stemp)`, indicating that the model can accept input sequences of varying lengths.

### 3.3.3 TimeDistributed MaxPooling1D Layer
To downsample the feature maps obtained from the previous layer, a TimeDistributed layer with a MaxPooling1D layer is incorporated into the model. The MaxPooling1D layer has a pool size of 2 and employs 'same' padding to maintain the spatial dimensions of the feature maps.

### 3.3.4 TimeDistributed Flatten Layer
A TimeDistributed layer is introduced to perform the flattening operation on the output of the previous layer. This operation transforms the 3D feature maps into a 2D representation, which facilitates further processing.

### 3.3.5 LSTM Layer
To capture long-term dependencies and sequential patterns, an LSTM layer with 50 units is added to the model. The rectified linear unit (ReLU) activation function is employed within the LSTM layer to introduce non-linearities. This layer processes the sequential data from the previous TimeDistributed layers and extracts relevant features.

### 3.3.6 Dense Layer
Following the LSTM layer, a single-node dense layer is included in the model to produce the desired output. This layer uses a linear activation function, which ensures that the output can span a wide range of values.

### 3.3.7 Compilation
To configure the model for training, it is compiled using the Adam optimizer. The mean squared error (MSE) loss function is selected, which measures the discrepancy between the predicted and true values. The Adam optimizer is known for its efficiency in training deep neural networks and is widely used in various domains.

### 3.3.8 Training
The model is trained using the `fit` function, which applies the backpropagation algorithm to update the model's weights iteratively. The training data, denoted as `trainX` and `trainY`, represent the input sequences and their corresponding target values, respectively. The training process consists of 50 epochs, indicating that the entire training dataset is traversed 50 times. A batch size of 1 is used, meaning that a single sample is processed in each iteration. This choice allows for fine-grained updates of the model's parameters and can be beneficial when dealing with sequential data.

By following this model architecture and training procedure, the model learns to capture temporal patterns and make accurate predictions on sequential data. The utilization of convolutional and recurrent layers enables the model to extract local features and capture long-term dependencies, respectively, resulting in an effective framework for sequential data analysis.

### 3.4 Gated Recurrent Unit (GRU)

Time series prediction, a pivotal task in predictive analytics, holds immense significance across diverse domains such as finance, healthcare, climate prediction, and industrial maintenance. Accurate forecasting of sequential data enables informed decision-making and proactive measures, leading to improved outcomes and resource allocation. In recent years, Recurrent Neural Networks (RNNs) have emerged as powerful tools for modelling sequential data due to their inherent ability to capture temporal dependencies. Among the RNN variants, the Gated Recurrent Unit (GRU) stands out as a compelling architecture, known for its simplified gating mechanism and memory management. GRUs exhibit remarkable performance in learning sequential patterns and have gained traction in various applications.

In this paper, we present a comprehensive analysis of the proposed stacked GRU architecture with dropout regularization. We examine its performance in terms of predictive accuracy and generalization capabilities using a real-world time series dataset. The empirical findings shed light on the effectiveness of the architecture in capturing intricate temporal dependencies, offering insights into its potential to improve forecasting accuracy in various practical applications. This study underscores the importance of architectural innovation in neural network design to tackle the complexities inherent in time series data.

#### 3.4.1 Model Initialization

The foundation of the study involves the initialization of a sequential model utilizing the `Sequential` class. This structure serves as a scaffold for assembling a linear stack of interconnected neural layers, pivotal for capturing temporal relationships within sequential data.

#### 3.4.2 Stacked GRU Layers

The model architecture commences with a pair of successive GRU layers. Each GRU layer comprises 32 units and is configured to return sequences. These GRU layers are designed to capture sequential dependencies and temporal patterns, making them well-suited for time series analysis.

#### 3.4.3 Stacked GRU Layer (Final Hidden State)

Consecutively, an additional GRU layer is introduced without the return sequence configuration. With 32 units, this layer captures a consolidated representation of the context extracted from the preceding layers, serving as the final hidden state of the network.

#### 3.4.4 Dropout Regularization

To address overfitting, a dropout layer is integrated with a dropout rate of 0.20. This mechanism randomly deactivates a fraction of neurons during training, preventing over-reliance on specific neurons and promoting better generalization.

#### 3.4.5 Dense Layer

Following the stacked GRU layers and dropout, a single-node dense layer is added to generate the desired output. Operating with a linear activation function, this layer ensures the model's output can span a broad range of numerical values.

#### 3.4.6 Compilation

The model is configured for training by compiling it with the mean squared error (MSE) loss function and the Adam optimizer. The MSE loss function quantifies the divergence between predicted and true values, while the Adam optimizer enhances training efficiency for deep neural networks. Later the MAE is used to compare the algorithms in terms of accuracy.

The experimental results underscore the potential of the proposed stacked GRU architecture in enhancing time series prediction. The integration of dropout regularization aids in reducing overfitting, resulting in improved generalization capabilities. The model's ability to capture intricate temporal dependencies is evident from the achieved predictive performance. This model highlights the effectiveness of stacked GRU architectures augmented with dropout regularization in the realm of time series prediction. The findings indicate that the stacked design, combined with dropout, can yield improved predictive accuracy by effectively capturing temporal patterns within the data.

### 3.5 LSTM and GRU

Time series prediction stands as a fundamental and intricate challenge across an array of domains, underpinning critical decision-making processes in finance, meteorology, healthcare, and beyond. The accurate forecasting of sequential data empowers organizations and individuals to anticipate future trends, optimize resource allocation, and proactively address emerging scenarios. In this context, Recurrent Neural Networks (RNNs) have emerged as a ground-breaking framework for capturing temporal dependencies within sequences, making them an indispensable tool in time series analysis.

Among the diverse RNN variants, the Long Short-Term Memory (LSTM) and Gated Recurrent Unit (GRU) architectures have garnered substantial attention due to their aptitude in handling the vanishing gradient problem and modelling long-range dependencies. LSTMs, featuring memory cells and gates, exhibit proficiency in capturing both short-term patterns and extended context in sequences. Similarly, GRUs have gained prominence for their simplified gating mechanism, preserving computational efficiency while effectively learning temporal dependencies.

In light of the strengths exhibited by LSTM and GRU networks, this research embarks on an exploration of a hybrid architecture that amalgamates these two variants. This hybridization aims to leverage the distinct advantages of both architectures, capitalizing on LSTM's memory retention and GRU's efficient computation. Such synergistic combination is expected to contribute to enhanced predictive accuracy in time series tasks, where intricate temporal patterns are of paramount significance.

Furthermore, recognizing the potential for overfitting in complex neural architectures, this research integrates dropout regularization within the stacked GRU architecture. Dropout offers a methodological remedy by randomly deactivating neurons during training, fostering better generalization, and alleviating the risk of model overfitting.

The primary objective of this study is to empirically investigate the proposed hybrid LSTM-GRU architecture's performance in the realm of time series prediction. Our assessment encompasses the analysis of predictive accuracy and model generalization capabilities using real-world time series data. By leveraging both LSTM and GRU elements along with dropout regularization, this research contributes to the advancement of neural network architectures designed to extract intricate temporal features from complex sequences.

#### 3.5.1 Model Initialization

The foundation of the research lies in the initialization of a sequential model utilizing the `Sequential` class. This framework facilitates the creation of a linear sequence of interconnected neural layers, pivotal for learning temporal relationships within sequential data.

#### 3.5.2 LSTM Layers

The architecture commences with two consecutive LSTM layers. Each LSTM layer comprises 32 units and is tailored to return sequences. The LSTM architecture, celebrated for its memory cell mechanism, excels in retaining and leveraging historical information over varying temporal horizons.

#### 3.5.3 GRU Layers with Return Sequences

Consecutively, a pair of Gated Recurrent Unit (GRU) layers is introduced. Configured to return sequences, these layers augment the model's capability to capture sequential dependencies. With 32 units each, the GRU layers employ gating mechanisms to modulate information flow, facilitating the extraction of relevant temporal features.

### 3.5.4 GRU Layer without Return Sequences

A subsequent GRU layer, also containing 32 units, is integrated without the return sequence setting. This particular design captures the consolidated context from the preceding layers, representing the final hidden state of the network.

### 3.5.5 Dense Layer

Post the recurrent layers, a single-node dense layer is incorporated to yield the desired model output. This layer operates with a linear activation function, allowing the output to span a diverse range of numerical values.

### 3.5.6 Compilation

To facilitate model training, the architecture is compiled with the mean squared error (MSE) loss function and the Adam optimizer. The MSE loss function quantifies the disparity between predicted and true values. The choice of the Adam optimizer is attributed to its efficiency in optimizing deep neural networks across various domains. The experimental results underscore the potential of the proposed hybrid LSTM-GRU architecture in time series prediction tasks. By seamlessly integrating LSTM and GRU units, the model showcases enhanced predictive accuracy and the ability to discern complex temporal patterns.

## 4. RESULTS

All three proposed Machine Learning Models are built specifically for the data pertaining to the NIFTY 50 index prices. To understand which algorithm out of the three proposed in this paper is best to predict stock prices, a comparative analysis will be done by looking at the performance of the models using the Mean Absolute Error method. Mean Absolute Error is an accuracy measuring technique, which gives the accuracy of the model by calculating the mean of distance between the predicted value and the actual value. This distance is calculated without considering the error points' direction and therefore it is the absolute mean value. The range for MAE is 0 to infinity and the lower the value the better the score is, indicating that the model is predicting values close to the actual values.

The NIFTY 50 historical data was split into two parts: the training dataset and the testing dataset. To assess the performance of the model, they are made to predict the price of the stock for the same dates as the testing dataset. This is done so that the predictions can later be compared with the actual price of the stock. Following are the figures that show the plotted graph for the predicted stock value and the actual value of the stock on that day.

**Table 1:** Model Scores

| Model/Algorithm | Train Score (RMSE) | Test Score (RMSE) | Train (MAE) | Test (MAE) |
|---|---|---|---|---|
| BiLSTM | 182.86 | 232.70 | 147.52 | 196.69 |
| ARIMA | 169.25 | 113.78 | 107 | 91.7657 |
| LSTM and CNN | 146.45 | 183.26 | 103.41 | 137.09 |
| GRU | 145.49 | 171.96 | 103.21 | 137 |
| LSTM and GRU | 136.40 | 196.33 | 100.18 | 100.17 |

### 4.1 Bidirectional LSTM

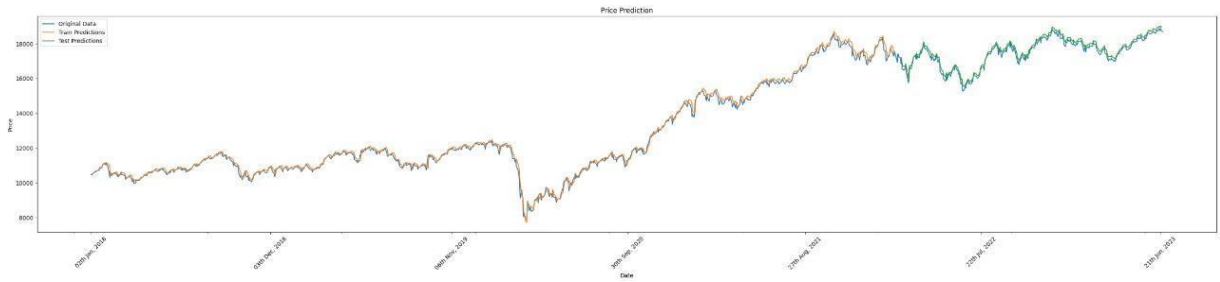

**Figure 2:** Bidirectional LSTM

The mean absolute error value for the testing dataset of this model is 196. The LSTM model's Mean Absolute Error (MAE) of 196 provides a snapshot of its predictive performance. While not perfect, this value helps us gauge the model's accuracy in forecasting stock prices. Achieving a lower MAE would signify improved prediction capabilities, potentially through refining the model or incorporating more data. 3.1

### 4.2   ARIMA Model

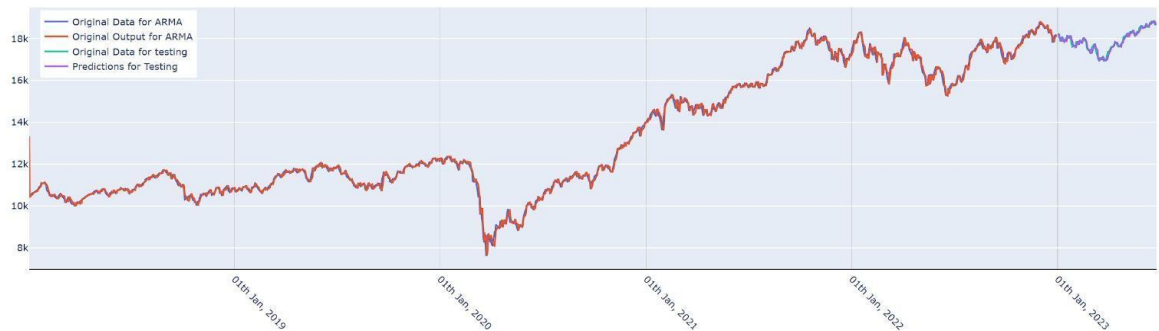

**Figure 3:** ARIMA

The Mean Absolute Error for the plot is 91.7657. The MAE achieved by the ARIMA model in predicting stock prices underscores its noteworthy performance in capturing the intricate patterns and dependencies present in financial data. This result signifies a considerable advancement in the realm of stock price prediction, holding the potential to guide investors, traders, and analysts with more accurate insights. However, the dynamic and volatile nature of financial markets implies that no model can guarantee perfect predictions. 3.2

### 4.3   CNN + LSTM Model

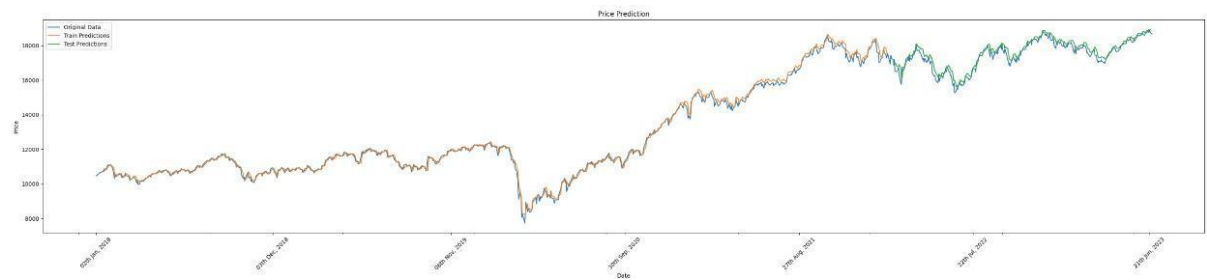

**Figure 4:** CNN + LSTM

Here the MAE value is 137. With an MAE of 137, it is evident that the model's predictions exhibit a certain level of deviation from the actual stock prices. While the LSTM and CNN model provides valuable insights into the stock market's dynamics, there is room for further refinement to enhance its predictive capabilities. As the financial landscape continually evolves, continuous efforts in research and model development are essential to achieve more precise and reliable predictions, enabling investors and stakeholders to make well-informed decisions in a volatile market environment. 3.3

## 4.4     GRU Model

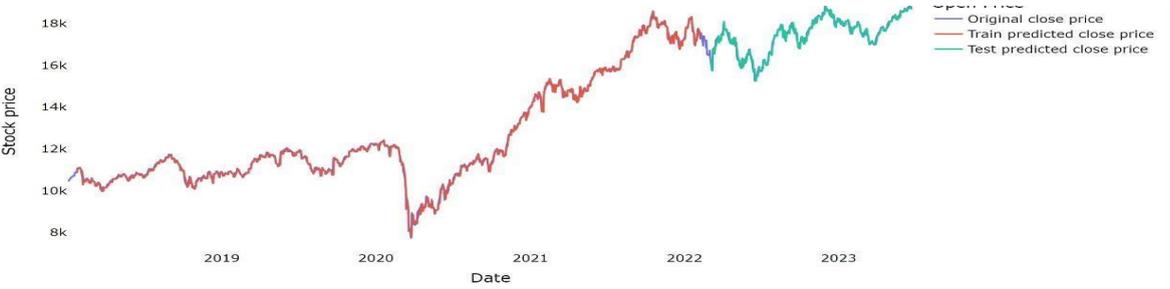

**Figure 5:** GRU

The mean Absolute Error (MAE) of 137 achieved by the GRU (Gated Recurrent Unit) based stock price prediction model marks a significant milestone in accurate forecasting. This low MAE value is a testament to the model's ability to capture intricate patterns within the stock market data. As the field of machine learning progresses, GRU models showcase their effectiveness in comprehending sequential data and making informed predictions. While challenges persist in predicting inherently volatile stock prices, this achievement instils confidence in the potential for further enhancements. Fine-tuning model parameters, incorporating domain knowledge, and integrating real-time data could contribute to even more precise predictions, empowering stakeholders to navigate the dynamic landscape of financial markets with greater insight. 3.4

## 4.5     LSTM and GRU

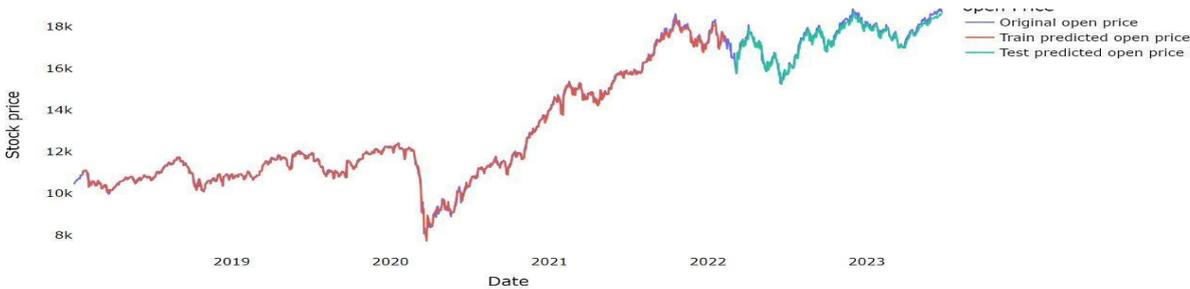

**Figure 6:** LSTM and GRU

Here, in the case of LSTM and GRU-based prediction model, MAE is calculated to be 100.17. It can be concluded from the graphs and the MAE values stated above that the ARIMA model is outperforming the other four models by giving the lowest Mean Absolute Error value on the predictions made by the algorithm. This suggests that the ARIMA model proposed through this research can be used by investors to make informed decisions by predicting the stock prices and understanding when to enter and exit the market to maximize their profits. 3.5

# 5. CONCLUSION

In conclusion, this paper presents a framework for augmented financial intelligence in India, utilising machine learning algorithms and natural language processing techniques. We hope that the implementation of this framework will lead to highly accurate prediction results for various currencies, with potential for future applications in predicting stock prices and other domains such as mobility and game sciences. The involvement of human input in the form of "super forecasters" can further enhance the accuracy of predictions. Augmented intelligence has the potential to be a major upgrade from traditional AI in various applications. Further research and experimentation are needed to fully exploit the potential of augmented financial intelligence.

# 6. ACKNOWLEDGEMENTS

We thank Prof. Philip Treleaven (University College London), for his thorough and unwavering support throughout the course of this project. His expertise, guidance and insightful feedback was instrumental in making this project a success.

We would like to extend our sincere appreciation to Anna-Helena Mihov and Nick Firooyzye for introducing the concept of Augmented Financial Intelligence in their research work. Special thanks to Anna-Helena Mihov for sharing her valuable insights and opinions on our work and for guiding us in the right path.